\documentclass[sigconf]{acmart}
\usepackage{graphicx}
\usepackage{textcomp}
\usepackage{xcolor}
\usepackage{booktabs}
\usepackage{multirow}
\usepackage{makecell}
\usepackage{bm}
\usepackage[linesnumbered,ruled,vlined]{algorithm2e}

\newcommand\modelName{MuSiC}

\usepackage{caption}
\usepackage{subcaption}
\usepackage[labelfont=bf,labelsep=colon]{caption} 

\AtBeginDocument{%
  }

\acmISBN{978-1-4503-XXXX-X/2018/06}

\begin{document}

\title{Leveraging Multimodal Data and Side Users for Diffusion  Cross-Domain Recommendation}


\author{Fan Zhang}
\email{zhang\_fan@bupt.edu.cn}
\affiliation{%
  \institution{School of Computer Science (National Pilot Software Engineering School), Beijing University of Posts and Telecommunications}
  \city{Beijing}
  \country{China}
}

\authornote{Also with Key Laboratory of Trustworthy Distributed Computing and Service (BUPT), Ministry of Education.}

\author{Jinpeng Chen}
\email{jpchen@bupt.edu.cn}
\affiliation{%
  \institution{School of Computer Science (National Pilot Software Engineering School), Beijing University of Posts and Telecommunications}
  \city{Beijing}
  \country{China}
}

\authornote{{Corresponding author.}}
\authornotemark[1]

\author{Huan Li}
\email{lihuan.cs@zju.edu.cn}
\affiliation{%
  \institution{The State Key Laboratory of Blockchain and Data Security, Zhejiang University}
  \city{Hangzhou}
  \country{China}
}

\author{Senzhang Wang}
\email{szwang@csu.edu.cn}
\affiliation{%
  \institution{Central South University}
  \city{Changsha}
  \country{China}
}

\author{Yuan Cao}
\email{e1124923@u.nus.edu}
\affiliation{%
  \institution{National University of Singapore}
  \city{Singapore}
  \country{Singapore}
}
\author{Kaimin Wei}
\email{kaiminwei@jnu.edu.cn}
\affiliation{%
  \institution{Jinan University}
  \city{Guangzhou}
  \country{China}
}

\author{Jianxiang He}
\email{hjx812143280@bupt.edu.cn}
\affiliation{%
  \institution{School of Computer Science (National Pilot Software Engineering School), Beijing University of Posts and Telecommunications}
  \city{Beijing}
  \country{China}
}
\authornotemark[1]

\author{Feifei Kou}
\email{koufeifei000@bupt.edu.cn}
\affiliation{%
  \institution{School of Computer Science (National Pilot Software Engineering School), Beijing University of Posts and Telecommunications}
  \city{Beijing}
  \country{China}
}
\authornotemark[1]

\author{Jinqing Wang}
\email{1439935227@qq.com}
\affiliation{%
  \institution{School of Computer Science (National Pilot Software Engineering School), Beijing University of Posts and Telecommunications}
  \city{Beijing}
  \country{China}
}
\authornotemark[1]

\renewcommand{\shortauthors}{Trovato et al.}

\begin{abstract}
Cross-domain recommendation (CDR) aims to address the persistent cold-start problem in Recommender Systems. Current CDR research concentrates on transferring cold-start users' information from the auxiliary domain to the target domain. 
However, these systems face two main issues: the underutilization of multimodal data, which hinders effective cross-domain alignment, and the neglect of side users who interact solely within the target domain, leading to inadequate learning of the target domain's vector space distribution.
To address these issues, we propose a model leveraging \underline{Mu}ltimodal data and \underline{Si}de users for diffusion \underline{C}ross-domain recommendation (MuSiC). 
We first employ a multimodal large language model to extract item multimodal features and leverage a large language model to uncover user features using prompt learning without fine-tuning.
Secondly, we propose the cross-domain diffusion module to learn the generation of feature vectors in the target domain. This approach involves learning feature distribution from side users and understanding the patterns in cross-domain transformation through overlapping users. 
Subsequently, the trained diffusion module is used to generate feature vectors for cold-start users in the target domain, enabling the completion of cross-domain recommendation tasks. Finally, our experimental evaluation of the Amazon dataset confirms that MuSiC achieves state-of-the-art performance, significantly outperforming all selected baselines. Our code is available: https://anonymous.4open.science/r/MuSiC-310A/.
\end{abstract}



\received{20 February 2007}
\received[revised]{12 March 2009}
\received[accepted]{5 June 2009}

\maketitle

\section{Introduction}
Recommendation systems \cite{wu2023instant, deng2024task} are widely used in daily applications, analyzing historical interaction data to learn user preferences and provide personalized recommendations.
However, most recommendation systems often struggle with cold-start users who lack sufficient interaction history, resulting in suboptimal recommendations. 
Cross-domain recommendation (CDR) \cite{SunGHDLCM23, yang2024not, guo2024prompt, song2024mitigating, ChenZLLJLLW24} helps mitigate this issue by transferring knowledge from an auxiliary domain, thereby improving personalization for cold-start scenarios.

Most CDR methods derive feature vectors based on interaction records within each domain. This approach leads to domain-specific feature vectors that hinder effective cross-domain transfer \cite{GongDSSLZ23}.
Existing studies align features based on user interests \cite{ZhaoZHZ023, ZhaoZLHWF24}, but they have not fully exploited multimodal data's inherent cross-domain universal properties, such as text and images.
As for methods using multimodal data, they mainly rely on 
visual and text encoders such as Pre-trained Language Models (PLMs) \cite{YangYL23, TangYW23}, which fail to leverage the powerful language understanding and generalization capabilities of large language models (LLMs). Unlike PLMs, typically trained with masked language modeling tasks, LLMs are trained using next-word prediction objectives. As discussed in \cite{wang2024llm4msr}, this fundamental difference enables LLMs to exhibit superior reasoning abilities and knowledge generation capabilities, such as analyzing user preferences, making them more suitable for complex multimodal tasks.
Building on this, we employ multimodal large language models (MLLMs) to integrate various data modalities and obtain domain-unified encoded representations, thereby achieving preliminary cross-domain alignment.

After obtaining user features, current CDR methods \cite{ManSJC17, ZhuTLZXZLH22} typically employ mapping methods for cross-domain transfer. However, existing mapping-based approaches often concentrate on precisely transferring features for specific data samples (e.g., overlapping users), which can lead to overfitting and limit the model's ability to generalize to unseen data. A more effective CDR model would be one that can accurately estimate the distribution of the target domain’s embedding space, thereby improving recommendation performance.
In recent years, diffusion models \cite{HoJA20} have seen significant advancements in the field of computer vision. The essence of using diffusion models for text-to-image generation lies in converting data from the textual domain into the image domain. Inspired by this, we treat auxiliary domain feature vectors as 'text' and target domain vectors as 'images'. This approach enables the generation of target domain feature vectors guided by information from the auxiliary domain.
Diffusion models effectively estimate the distribution of the target domain’s embedding space by adapting to its features. By adding noise to the feature vectors from the auxiliary domain and gradually removing them, the model incrementally learns the target domain's characteristics, refining the vectors into representations that align more closely with the target domain.

\begin{figure}[h]
  \centering
  \begin{subfigure}[b]{0.45\textwidth}
    \centering
    \includegraphics[width=\textwidth]{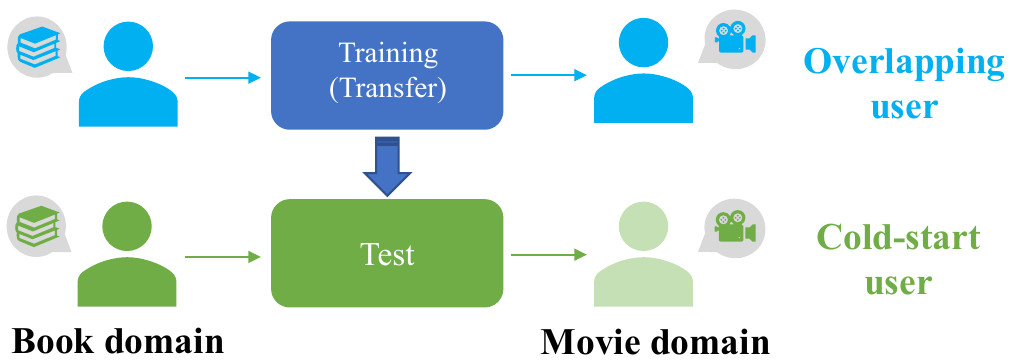}
    \caption{Existing CDR methods}
    \label{fig1a}
  \end{subfigure}

  \begin{subfigure}[b]{0.45\textwidth}
    \centering
    \includegraphics[width=\textwidth]{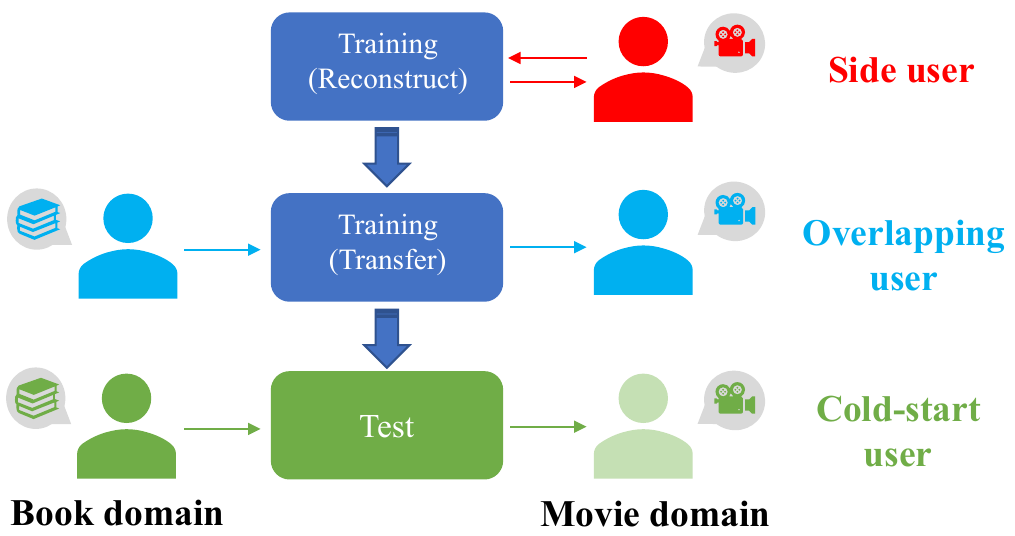}
    \caption{Our work}
    \label{fig1b}
  \end{subfigure}
  \caption{(a) Existing CDR methods typically learn cross-domain transfers by overlapping users. (b) Our approach first learns the feature distribution of the target domain by reconstructing side users and then leverages overlapping users to facilitate cross-domain transfer.}
  \label{fig1}
\end{figure}

Accurately estimating the distribution of the target domain's embedding space requires sufficient sample data. However, existing CDR methods primarily rely on overlapping users to learn about the target domain (as shown in Figure~\ref{fig1a}), failing to fully leverage side users who interact exclusively within the target domain.
As illustrated in Figure~\ref{fig1b}, in practical scenarios, there are three types of users in the target domain: overlapping users who interact in both domains, cold-start users who require predictions, and side users who interact only in the target domain. 
Side users are crucial for improving CDR performance. By reconstructing their feature vectors, we can better understand and adapt to the target domain's feature distribution and user preferences.

To address the above problems, this paper introduces a model leveraging \textbf{Mu}ltimodal data and \textbf{Si}de users for diffusion \textbf{C}ross-domain recommendation (\modelName). This model utilizes multimodal data for preliminary cross-domain alignment and learns target domain representations from side users. Firstly, \modelName~employs MLLMs to acquire high-precision feature vectors from the multimodal data using prompt learning. These content-based vectors facilitate preliminary cross-domain alignment. 
Subsequently, \modelName~proposes a cross-domain diffusion module with a two-stage process to generate target domain feature vectors. In the first stage, the model leverages side user data to capture the target domain's feature distribution. In the second stage, it transfers features from the auxiliary domain to the target domain using overlapping user data. This two-stage approach allows the model to effectively learn the target domain’s feature distribution and transfer domain-specific knowledge.
Finally, \modelName~generates target domain feature vectors for cold-start users and make cross-domain recommendations by calculating the dot product between the cold-start users and the items in the target domain.
Additional experiments conducted in dual cold-start scenarios (recommending cold-start items to cold-start users) demonstrate the effectiveness of the \modelName~method.

Our contributions are summarized as follows:

\begin{itemize} 
 \item We leverage MLLMs to effectively extract features from multimodal data, addressing the underutilization of such data and enhancing cross-domain alignment.

 \item We propose a two-stage cross-domain diffusion module, which utilizes side users to capture the feature distribution of the target domain and facilitate effective cross-domain knowledge transfer.
 
 \item We conduct extensive experiments on multiple Amazon datasets, and the results demonstrate that \modelName~significantly improves the accuracy of cross-domain recommendations, particularly for cold-start items.
\end{itemize}

\section{Related work}

\textbf{Cross-domain Recommendation.} 
Cross-domain recommendation \cite{xu2023neural, zhang2023cross, an2024ddcdr, chen2024improving} typically involves transferring feature vectors across domains via mapping techniques.  
CMF \cite{SinghG08} integrates knowledge across domains by linking multiple rating matrices.  
EMCDR \cite{ManSJC17} learns a mapping function to infer latent features for cold-start users in the target domain.  
SSCDR \cite{KangHLY19} uses a cross-domain mapping function to encode distance information from both labeled and unlabeled data.  
LACDR \cite{WangZZWZH21} employs an encoder-decoder structure to create a mapping function that aligns low-dimensional embedding spaces across domains.  
PTUPCDR \cite{ZhuTLZXZLH22} trains a meta-network to generate personalized mapping functions based on various user preferences.  
CATN \cite{ZhaoLXDS20} and RC-DFM \cite{FuPWXL19} use user reviews to extract aspect-level information for cross-domain recommendations.  
OmniMatch \cite{DaiEAT25} generates auxiliary reviews for cold-start users and employs domain adversarial training and supervised contrastive learning to ensure domain-invariant features.  
The structure of these methods is illustrated in Figure~\ref{fig1a}. However, they often overlook the information from side users in the target domain.  
In contrast, our method first aligns cross-domain information using multimodal semantic data, then incorporates side user information to learn the feature distribution in the target domain, enhancing recommendation performance.

\begin{figure*}[h]
 \centering
 \includegraphics[width=0.99\textwidth]{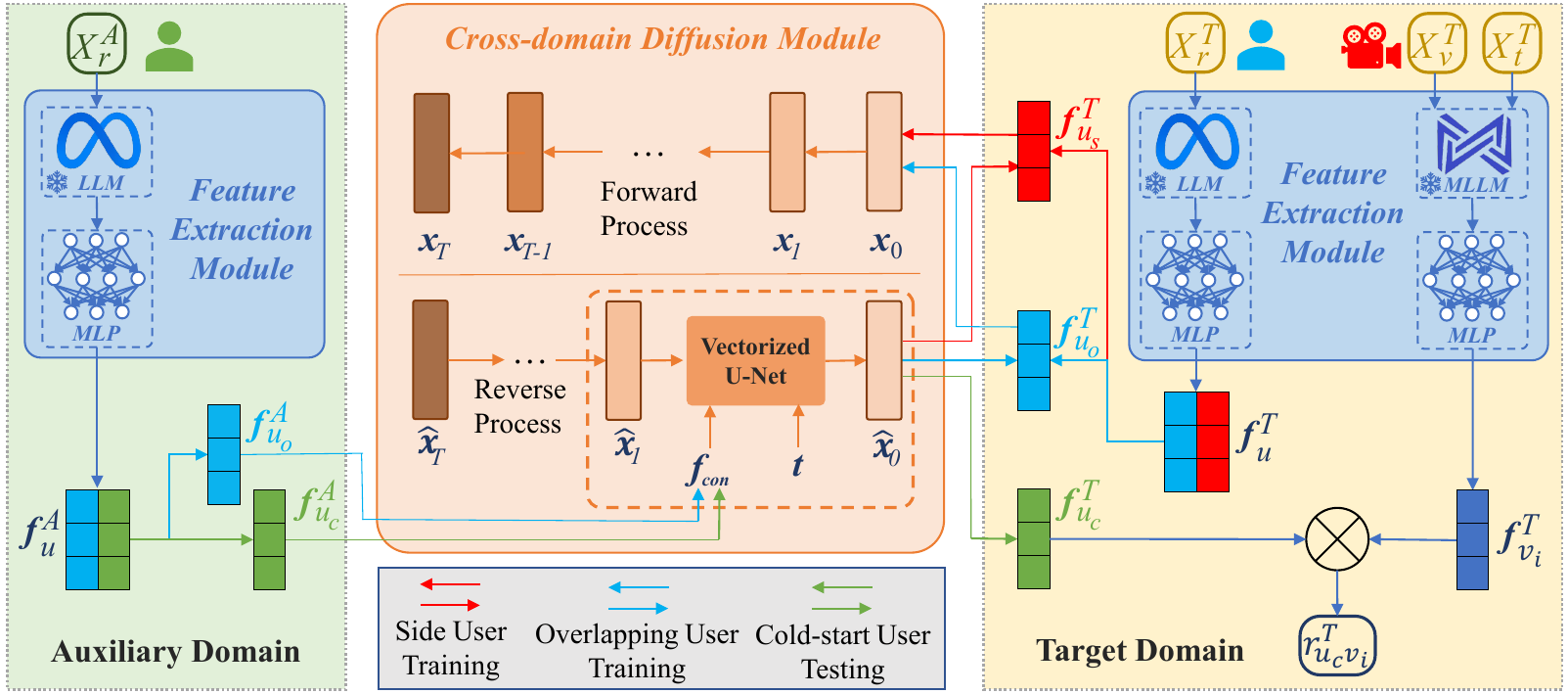}
 \caption{Overall architecture of \modelName.
Firstly, we utilize the user feature extraction module in the auxiliary and target domains to obtain features for side, overlapping, and cold-start users. 
Then, the features of side and overlapping users are used to train the cross-domain diffusion module, designed to generate feature vectors for cold-start users in the target domain.
Finally, these generated feature vectors are combined with item features extracted by the item feature extraction module in the target domain to produce recommendations.}
 \label{fig2}
\end{figure*}

\textbf{Multimodal-based Recommendation.}
With the growing abundance of multimodal data, many researchers have explored its potential to alleviate data sparsity in recommender systems \cite{yang2024multimodal, li2024attribute, shang2024improving, fu2024iisan, guo2024lgmrec}. 
MMGCN \cite{wei2019mmgcn} constructed a user-item bipartite graph to learn modality-specific representations, which are fused into a unified representation to enhance recommendation accuracy. 
GRCN \cite{wei2020graph} leveraged multimodal content to refine the interaction graph structure, mitigating the effects of false-positive edges on recommendation performance.
Some cross-domain recommendation work also explores multimodal scenarios. 
MOTKD \cite{YangYL23} used graph attention networks to obtain representations of different modalities and applied knowledge distillation to facilitate cross-domain knowledge transfer for overlapping users. 
P2M2-CDR \cite{wang2024privacy} proposes a privacy-preserving framework that improves cross-domain recommendation for overlapping users using multimodal data and disentangled embeddings, while ensuring privacy with local differential privacy.
The aforementioned approaches leverage multimodal data to enhance cross-domain recommendations for overlapping users but fail to address the cold-start user problem. In contrast, our approach focuses on cold-start users, enabling recommendations even without interactions in the target domain.

\textbf{Diffusion-based Recommendation.}
In recent years, the improved generative capabilities of diffusion models have led to a growing number of studies in the recommendation domain \cite{MaYMXM24, ZhaoWXSFC24, wang2024diff, long2024diffusion, chen2024link}.  DiffRec \cite{WangXFL0C23} was the first to employ diffusion models for recommendations, introducing a denoising method to learn user-item interactions.  
Diff4Rec \cite{Wu0CLHS023} improves performance in sparse data scenarios by corrupting and reconstructing user-item interactions.  
Diff-POI \cite{qin2023diffusion} used the diffusion process to sample user spatial preferences for next point-of-interest recommendations.  
DiffKG \cite{jiang2024diffkg} integrated a generative diffusion model with data augmentation for robust knowledge graph representation learning.  
DiffMM \cite{jiang2024diffmm} combines a modality-aware graph diffusion model with cross-modal contrastive learning to improve user representation.  
DiffCDR \cite{abs-2402-02182} is the only existing work that applies diffusion models in cross-domain recommendations, generating user embeddings in the target domain using a diffusion process.  
In contrast to existing research, we utilize data from side users to enhance the accuracy of the embeddings generated by the diffusion models.

\section{METHODOLOGY}
\subsection{Problem Definition}
The CDR task involves an auxiliary domain and a target domain. In each domain, we have a user set $U = \{u_1, u_2, \ldots, u_i\}$, an item set $V = \{v_1, v_2, \ldots, v_j\}$, and a rating matrix $\bm{R}$. For each element of $\bm{R}$, $r_{ij}$ denotes the rating of item $v_j$ by $u_i$. Each item has multimodal content data, including the item's image data $X_v$ and the textual descriptions $X_t$. Each user has review data $X_r$. The superscripts $A$ and $T$ denote the auxiliary and target domains, respectively. The overlapping users are defined as $U_o$, the cold-start users as $U_c$, and the side users as $U_s$.
Therefore, the primary objective of this work is to learn a predictive function $f$, which takes a cold-start user $u_c \in U_c$ and a target domain item $v_i \in V^T$ as input, and outputs a predicted rating $r^{T}_{u_c,v_i}$. The goal is to minimize the difference between the predicted ratings and the ground-truth ratings.

Figure~\ref{fig2} shows the structure of our proposed \modelName. There are two main components of \modelName: (1) feature extraction module and (2) cross-domain diffusion module. 
The details are described below.

\subsection{Feature Extraction Module}

The feature extraction module includes item and user feature extraction. We first describe the item feature extraction in detail, and highlight the differences in user feature extraction later, as the approach is similar.

Given the powerful modeling capabilities of LLMs, we use the multimodal large language model MiniCPM-Llama3-V 2.5 (referred to as MiniCPM-V) to extract item feature vectors from multimodal data. Since MiniCPM-V is a chat-based model and cannot directly generate item feature vectors, we overcome this limitation by using fused encoding of multimodal inputs to represent item features.

We begin by applying prompt learning to the MiniCPM-V model, with images $X_v$ and text descriptions $X_t$ as input, guiding the model to summarize the key features of the item based on the combined image and text information. This allows the model to extract fused item features without fine-tuning.  
Next, we extract the hidden layer outputs and compute the average of the tokens from the first and last hidden layers \cite{sentence}, resulting in the first-last-avg representation, which serves as the multimodal item features.


  

\begin{equation}
\bm{h}_{first-last-avg} = \frac{1}{2} \left( \frac{1}{n} \sum_{i=1}^{n} \bm{H}_1^i + \frac{1}{n} \sum_{i=1}^{n} \bm{H}_L^i \right)
\end{equation}
where $\bm{H}_1$ and $\bm{H}_L$ represent the hidden layer outputs of the first and last layers respectively, and $n$ is the number of tokens.

After acquiring the encoded representation of the item multimodal features, we input it into the Multilayer Perceptron (MLP) using \textit{tanh} activation functions to extract the final item features $\bm{f}_{v}$.


Only textual review data $X_r$ is available for users, with no additional multimodal data. Therefore, we use the Llama3-8B model as an example to obtain their feature vectors $\bm{f}_{u}$. The remainder of the process follows the same steps to obtain the item feature vectors.

\subsection{Cross-domain Diffusion Module}
In this section, we explore the cross-domain diffusion module, designed to learn the generation pattern of target domain feature vectors to generate features for cold-start users in the target domain. This module comprises two main processes \cite{HoJA20}: the forward process and the reverse process. The forward process gradually adds Gaussian noise to the target domain feature vectors $\bm{f}_u^T$ to disrupt the original data distribution. Conversely, the reverse process entails progressively restoring the perturbed representations from a disordered state, ultimately restoring them to the target domain feature vectors.

\subsubsection{Forward Process}
Since the cross-domain diffusion module is designed to generate target domain feature vectors, we define both the original input $\bm{x}_0$ and the generated target $\hat{\bm{x}}_0$ as the feature vectors $\bm{f}_u^T$ of target domain users. The forward process constitutes a Markov chain, where Gaussian noise is gradually added to $\bm{x}_0$ as follows:
\begin{equation}
q(\bm{x}_{1:T} \mid \bm{x}_0) = \prod_{t=1}^{T} q(\bm{x}_t \mid \bm{x}_{t-1})
\end{equation}
\begin{equation}
q(\bm{x}_t \mid \bm{x}_{t-1}) = \mathcal{\bm{N}}\left( \bm{x}_t; \sqrt{1 - \beta_t} \bm{x}_{t-1}, \beta_t \mathbf{I} \right)
\end{equation}
where $t$ represents the diffusion step, and $\beta_t \in (0, 1)$ denotes the scale of the added Gaussian noise, which is determined by a linear schedule. By leveraging the reparameterization trick \cite{HoJA20} and the additivity of two independent Gaussian noises, we can obtain $\bm{x}_t$ at any time step. Hence, the forward process can be expressed as follows:
\begin{equation}
q(\bm{x}_t \mid \bm{x}_0) = \mathcal{\bm{N}}\left( \bm{x}_t; \sqrt{\bar{\alpha}_t} \bm{x}_t, (1 - \bar{\alpha}_t) \mathbf{I} \right)
\end{equation}
where $\alpha_t = 1 - \beta_t$, $\bar{\alpha}_t = \prod_{t'=1}^{t} \alpha_{t'}$, and then we can reparameterize
$\bm{x}_t = \sqrt{\bar{\alpha}_t} \bm{x}_0 + \sqrt{1 - \bar{\alpha}_t} \bm{\epsilon}$ with $\bm{\epsilon} \sim \mathcal{\bm{N}}(\bm{0}, \mathbf{I})$. 

\subsubsection{Reverse Process}
The reverse process of the diffusion model aims to iteratively denoise $\bm{x}_t$ over $T$ steps, ultimately approximating the target domain feature vector $\bm{x}_0$($\bm{f}_u^T$).

We use the auxiliary domain feature vectors of users $\bm{f}_u^A$ as additional conditions to guide the conditional generation of the project representation. 
It is worth noting that for side users, who lack historical records in the auxiliary domain, the generation of target domain feature vectors proceeds without conditional guidance. 
The conditional generation process at each step can be expressed as:
\begin{equation}
p_{\theta} (\bm{x}_{t-1} \mid \bm{x}_t) = \mathcal{\bm{N}} \left( \bm{x}_t; \mu_{\theta} (\bm{x}_t, t, \bm{f}_{con}), \Sigma_{\theta} (\bm{x}_t, t) \right)
\end{equation}
where $\mu_{\theta} (\bm{x}_t, t, \bm{f}_{con})$ and $\Sigma_{\theta} (\bm{x}_t, t)$ are the mean and covariance of the Gaussian distribution, respectively, and $\bm{f}_{con}$ represents the conditional guidance information. To maintain training stability and simplify calculations, we set $\Sigma_{\theta} (\bm{x}_t , t) = \sigma^2 (t)\mathbf{I}$ following \cite{HoJA20}. The mean $\mu_{\theta} (\bm{x}_t, t, \bm{f}_{con})$ can be calculated as follows:
\begin{equation}
\mu_{\theta} (\bm{x}_t, t, \bm{f}_{con}) = \frac{1}{\sqrt{\alpha_t}} \left( \bm{x}_t - \frac{\beta_t}{\sqrt{1 - \bar{\alpha}_t}} f_{\theta} (\bm{x}_t, t, \bm{f}_{con}) \right)
\end{equation}
where conditional estimator $f_{\theta} (\cdot)$ is typically constructed using various deep neural networks, such as MLP and U-net \cite{RonnebergerFB15}. 
In this context, we propose the vectorized U-Net to adapt the diffusion model for processing high-dimensional feature vectors instead of image data. 

\subsubsection{Vectorized U-Net}
The cross-domain diffusion module operates on vector representations $\bm{f}_u^T$ rather than image data, marking a significant departure from traditional U-Net-based diffusion models. By adapting the U-Net to vector-based operations, we extend the applicability of diffusion models to CDR scenarios involving multimodal data. 
\textbf{Initialization}  
The algorithm begins by computing the timestep embedding \( \bm{\bm{temb}} \), which encodes temporal information into a fixed-dimensional representation:
\begin{equation}
\bm{\text{TimestepEmbedding}}(t) = \left[ \sin(t \cdot \omega_i), \cos(t \cdot \omega_i) \right]_{i=1}^{d/2}
\end{equation}
where \( \omega_i = \exp\left(-\frac{\log(10000)}{d/2 - 1} \cdot i \right) \). 
If the optional conditional guidance vector \( \bm{f}_{con} \) is provided, it is added to the input feature vector \( x_t \). Otherwise, the input feature vector remains unchanged.

\textbf{Down-Sampling Stage}
The down-sampling stage transforms the input feature vector \( \bm{x}_t \) into a lower-dimensional latent representation while incorporating both temporal and conditional guidance. 

\textbf{Middle Block}
The middle block acts as the bottleneck of the framework, extracting the core features required for target domain representation. 

\textbf{Up-Sampling Stage} 
The up-sampling stage reconstructs the target domain feature vector \( \hat{\bm{x}}_{t-1} \) by progressively increasing the feature dimensionality and reintroducing hierarchical information.

\textbf{Output Layer} 
The predicted feature vector \( \hat{\bm{x}}_{t-1} \) is generated through a linear transformation, serving as an intermediate output that reconstructs the target for the current timestep. This iterative process continues until the final output \( \hat{\bm{x}}_0 \) is obtained.

\subsubsection{Inference}  
For a cold-start user \( u_c \in U_c \), the target domain feature vector is initialized with pure Gaussian noise. 
Through the iterative denoising process, the reverse diffusion process incrementally refines this feature vector, with the auxiliary domain feature vectors \(\bm{f}_{u_c}^A\) providing conditional guidance. After \( T \) steps, the reconstructed target domain feature vector \(\hat{\bm{x}}_0\) is obtained. This vector represents the cold-start user in the target, \(\bm{f}_{u_c}^T\), and is ready for downstream tasks such as recommendations.

This inference mechanism leverages knowledge from the auxiliary domain to generate accurate and contextually relevant representations for cold-start users in the target domain. By leveraging the cross-domain diffusion framework, the model ensures that the generated features align with the target domain's data distribution while retaining the essential characteristics of the cold-start users' behavior in the auxiliary domain.

\subsection{Cross-domain Recommendation}  
\subsubsection{Rating Function}  
For a cold-start user \( u_c \in U_c \) and an item \( v_i \in V^T \), the predicted rating \( r^T_{u_cv_i} \) is computed as:
\begin{equation}
r^T_{u_cv_i} = \bm{f}^T_{u_c} \cdot \bm{f}^T_{v_i}
\end{equation}
where \(\bm{f}^T_{u_c} \) and \(\bm{f}^T_{v_i} \) are the feature vectors of the user and item, respectively. This interaction score reflects the user's preference for the item.

\subsubsection{Top-N Recommendation}  
The predicted ratings \( r^T_{u_cv_i} \) are sorted in descending order for each cold-start user, and the top-N items with the highest scores are directly selected to form the recommendation list.
Unlike traditional methods that rely on explicit comparisons between positive and negative samples to optimize ranking, our approach derives the Top-N recommendations as a natural byproduct of the rating prediction process, facilitating the validation of the overall algorithm.

\begin{table*}[h]
\centering
\caption{Statistics of datasets in three tasks.}
\begin{tabular}{c|cc|ccc|cc|cc}
\toprule
\multirow{2}{*}{\textbf{Tasks}}& \multicolumn{2}{c|}{\textbf{Domain}} & \multicolumn{3}{c|}{\textbf{User}} & \multicolumn{2}{c|}{\textbf{Item}} & \multicolumn{2}{c}{\textbf{Rating}} \\
& \textbf{Auxiliary} & \textbf{Target} & \textbf{Auxiliary} & \textbf{Target}& \textbf{Overlap}  & \textbf{Auxiliary} & \textbf{Target} & \textbf{Auxiliary} & \textbf{Target} \\
\midrule
Task1 & {Movie} & {Music} & {18,045}& {10,292}& {2,165}& {156,884}& {159,997}& {881,003}& {513,247}\\
Task2 & {Book} & {Movie} & {53,442}& {18,045}& {3,545}& {674,378}& {156,884}& {2,772,516}& {881,003}\\
Task3 & {Book} & {Music} & {53,442}& {10,292}& {1,519}& {674,378}& {159,997}& {2,772,516}& {513,247}\\
\bottomrule
\end{tabular}

\label{table1}
\end{table*}

\subsection{Model Training}
The training of the~\modelName~is divided into two stages, each corresponding to different user groups: (1) feature alignment for side users, and (2) joint optimization of feature alignment and rating prediction for overlapping users.

\subsubsection{Loss Functions}

The diffusion model compels the posterior distribution to converge toward the prior distribution during the reverse process. This optimization is quantified using the KL divergence:
\begin{equation}
\mathcal{L}_{1} = D_{KL} \left( q(\bm{x}_{t-1} \mid \bm{x}_t, \bm{x}_0) \parallel p_{\theta} (\bm{x}_{t-1} \mid \bm{x}_t) \right)
\end{equation}

Based on the derivation in the DDPM \cite{HoJA20}, it can be simplified to a Mean-Squared Error (MSE) loss with the following formula:
\begin{equation}
\mathcal{L}_{dm} = \mathbb{E}_{\bm{x}_0, \bm{x}_t} \left[ \left\| \bm{x}_0 - f_{\theta} (\bm{x}_t, t, \bm{f}_{con}) \right\|^2 \right]
\label{eq1}
\end{equation}
where $\bm{x}_0$ denotes the target domain feature vector $\bm{f}_{u_o}^T$, and $f_{\theta}$ refers to the Vectorized U-Net to generate $\hat{\bm{x}}_0$.

\paragraph{Diffusion Loss for Side Users}  

For side users \( u_s \in U_s \), due to the absence of historical records in the auxiliary domain, the generation of target domain feature vectors $\bm{x}_0$($\bm{f}_{u_s}^T$) is performed without conditional guidance. The diffusion loss for side users is defined as:
\begin{equation}
\mathcal{L}_{dm\_u_s} = \mathbb{E}_{\bm{x}_0, \bm{x}_t} \left[ \left\| \bm{x}_0 - f_{\theta} (\bm{x}_t, t) \right\|^2 \right]
\end{equation}

\paragraph{Diffusion Loss for Overlapping Users}  
For overlapping users \( u_o \in U_o \), the diffusion loss \( \mathcal{L}_{dm\_u_o} \) is used to reconstruct the target domain feature vector $\bm{x}_0$($\bm{f}_{u_o}^T$) by leveraging the auxiliary domain feature \( \bm{f}^A_{u_o} \) as conditional guidance. The loss $\mathcal{L}_{dm\_u_o}$ is given by Equation \ref{eq1}.

\paragraph{Joint Loss for Overlapping Users}  
The joint loss for overlapping users combines \( \mathcal{L}_{dm\_u_o} \) and the rating prediction loss \( \mathcal{L}_{rating} \), with the two components balanced by a weighting parameter \(\lambda\), defined as follows:
\begin{equation}
\mathcal{L}_{rating} = \frac{1}{N} \sum_{u_o \in U_o} \sum_{v_i \in V^T} \left( \hat{r}_{u_ov_i} - r_{u_ov_i} \right)^2
\end{equation}
\begin{equation}
\mathcal{L}_{joint} = \lambda \mathcal{L}_{dm\_u_o} + (1 - \lambda) \mathcal{L}_{rating}
\end{equation}

\subsubsection{Optimization}  

The total loss for a single training iteration is:
\begin{equation}
\mathcal{L}_{total} = \mathcal{L}_{dm\_u_s} + \mathcal{L}_{joint}
\end{equation}
The training process is divided into two stages: the model first optimizes the diffusion loss \( \mathcal{L}_{dm\_u_s} \) for side users, and then minimizes the joint loss \( \mathcal{L}_{joint} \) for overlapping users. This sequential approach ensures effective alignment of feature distributions and accurate rating predictions.

\section{Experiment}

\begin{table*}[h]
\centering
\caption{Performance of cross-domain recommendation on MAE, RMSE, and NDCG. The best and second-best results are highlighted in boldface and underlined respectively.}
\begin{tabular}{c|c|c|ccccccccc|c}
\toprule
Tasks & $\beta$ & Metric & CMF & EMCDR & SSCDR & LACDR  & PTUPCDR & DiffCDR & CATN    &OmniMatch& \modelName~& Improve \\
\midrule
\multirow{9}{*}{Task 1} 
& \multirow{3}{*}{20\%} 
& MAE & 0.9468 & 0.7768 & 0.7954 & 0.7973 & 0.7766 & 0.7914 & \underline{0.7289} &0.7798& \textbf{0.6636}& 8.96\%\\
 &  & RMSE & 1.2350 & 1.0297 & 1.0548 & 1.0380 & 1.0629 & 1.0510 & \underline{1.0007} &1.0085& \textbf{0.9243}& 7.63\%\\
   &  & NDCG& 0.8271& 0.8390& 0.8393& 0.8387& 0.8366& 0.8388& \underline{0.8614} &0.8559& \textbf{0.8650}& 0.42\%\\
   \cline{2-13}
& \multirow{3}{*}{50\%} 
& MAE & 0.9614 & 0.7844 & 0.8089 & 0.7936 & 0.7847 & 0.7976 & \underline{0.7320} &0.8452& \textbf{0.6623}& 9.52\%\\
 &  & RMSE & 1.2647 & 1.0445 & 1.0655 & 1.0421 & 1.0644 & 1.0540 & \underline{1.0097} &1.0630& \textbf{0.9228}& 8.61\%\\
   &  & NDCG & 0.8260& 0.8349& 0.8357& 0.8360& 0.8330& 0.8349& \underline{0.8590} &0.8378& \textbf{0.8637}& 0.55\%\\
 \cline{2-13}
& \multirow{3}{*}{80\%} 
& MAE & 0.9739 & 0.7998 & 0.8072 & 0.7797 & 0.8112 & 0.7988 & \underline{0.7444} &0.9200& \textbf{0.6901}& 7.29\%\\
 &  & RMSE & 1.2673 & 1.0698 & 1.0876 & 1.0532 & 1.0885 & 1.0768 & \underline{1.0388} &1.1129& \textbf{0.9570}& 7.87\%\\
  &  & NDCG & 0.8299& 0.8390& 0.8381& 0.8386& 0.8381& 0.8394& \underline{0.8588} &0.8355& \textbf{0.8600}& 0.14\%\\
 \hline
\multirow{9}{*}{Task 2} 
& \multirow{3}{*}{20\%} 
& MAE & 0.9508 & 0.8313 & 0.8607 & 0.8286& 0.8462 & 0.8308 & 0.8413 &\underline{0.7995}&  \textbf{0.7646}& 4.36\%\\
 &  & RMSE & 1.2460 & 1.0874 & 1.1128 & 1.0823& 1.1105 & 1.0984 & 1.1098 &\underline{1.0313}& \textbf{1.0136}& 1.72\%\\
   &  & NDCG & 0.8081& 0.8338& 0.8333& 0.8359& 0.8294& 0.8359& 0.8419&\underline{0.8435}& \textbf{0.8543}& 1.28\%\\
 \cline{2-13}
& \multirow{3}{*}{50\%} 
& MAE & 0.9883 & 0.8626 & 0.8724 & 0.8574 & 0.8787 & 0.8849 & 0.8504&\underline{0.8326}& \textbf{0.7889}& 5.24\%\\
 &  & RMSE & 1.2859 & 1.1224 & 1.1422 & 1.1136& 1.1486 & 1.1293 & 1.1276 &\underline{1.0732}& \textbf{1.0450}& 6.16\%\\
   &  & NDCG & 0.7990& 0.8231& 0.8203& 0.8222& 0.8206& 0.8189& 0.8329&\underline{0.8342}& \textbf{0.8482}& 2.63\%\\
 \cline{2-13}
& \multirow{3}{*}{80\%} 
& MAE & 1.0180 & 0.8672 & 0.8856 & 0.8587 & 0.8850 & 0.8525 & \underline{0.8404} &0.8838&  \textbf{0.7954}& 5.35\%\\
 &  & RMSE & 1.3179 & 1.1357 & 1.1394 & 1.1130& 1.1581 & 1.1287 & 1.1222 &\underline{1.0941}& \textbf{1.0466}& 4.34\%\\
   &  & NDCG & 0.7941& 0.8198& 0.8207& 0.8205& 0.8144& 0.8200& \underline{0.8318} &0.8140& \textbf{0.8443}& 1.50\%\\
 \hline
 \multirow{9}{*}{Task 3} 
 & \multirow{3}{*}{20\%} 
& MAE & 0.8987 & 0.7722 & 0.7762 & 0.7536 & 0.7705 & 0.7764 & \underline{0.7213} &0.7954& \textbf{0.6235}& 13.56\%\\
 &  & RMSE & 1.1836 & 1.0317 & 1.0428 & 1.0169 & 1.0356 & 1.0345 & \underline{1.0030} &1.0061& \textbf{0.8654}& 13.72\%\\
   &  & NDCG & 0.8359& 0.8466& 0.8454& 0.8451& 0.8460& 0.8421& \underline{0.8679} &0.8673& \textbf{0.8713}& 0.39\%\\
 \cline{2-13}
& \multirow{3}{*}{50\%} 
& MAE & 0.9207 & 0.7977 & 0.7853 & 0.7765 & 0.7825 & 0.8561 & \underline{0.7281} &0.8160& \textbf{0.6696}& 8.03\%\\
 &  & RMSE & 1.2074 & 1.0639 & 1.0578 & 1.0400 & 1.0525 & 1.0893 & \underline{1.0156} &1.0256& \textbf{0.9271}& 8.71\%\\
   &  & NDCG & 0.8318& 0.8410& 0.8409& 0.8411& 0.8383& 0.8396& \underline{0.8650} &0.8485& \textbf{0.8681}& 0.36\%\\
 \cline{2-13}
 & \multirow{3}{*}{80\%} 
& MAE & 0.9451 & 0.8211 & 0.7954 & 0.7853 & 0.8207 & 0.7904 &\underline{ 0.7306} &0.8565& \textbf{0.6899}& 5.57\%\\
 &  & RMSE & 1.2204 & 1.0783 & 1.0626 & 1.0426 & 1.0987 & 1.0637 & \underline{1.0169} &1.0431& \textbf{0.9416}& 7.40\%\\
   &  & NDCG & 0.8269& 0.8369& 0.8367& 0.8371& 0.8354& 0.8366& \textbf{0.8616}&0.8346& \underline{0.8546}& -0.81\%\\
 
\bottomrule
\end{tabular}

\label{table2}
\end{table*}

In this section, we conducted extensive experiments across three cross-domain tasks to evaluate our \modelName~method. 
\subsection{Experimental Settings}
\subsubsection{Dataset} This study uses the Amazon dataset \cite{hou2024bridging}, which includes 21 categories. We follow the method in \cite{ZhuTLZXZLH22} and use three subsets: movies, music, and books, to create three cross-domain experiments: movie-music, movie-book, and book-music. The dataset includes item metadata and user reviews. For item representation, we select the "title", "description", and "categories" from the metadata for the text modality, and the first image from the "images" for the imaging modality. For users, we use their review texts as input for the text modality. We filter out reviews shorter than 10 words, items without descriptions or images, and users with fewer than 20 reviews. The dataset statistics are shown in Table~\ref{table1}.

\subsubsection{Metrics.} The Amazon dataset includes user ratings for items, ranging from 0 to 5. Following \cite{ZhuTLZXZLH22}, we use Mean Absolute Error (MAE) and Root Mean Square Error (RMSE) to evaluate prediction accuracy. Additionally, we include Normalized Discounted Cumulative Gain (NDCG@20) to assess the ranking quality of the top 20 recommended items.

\subsubsection{Baselines.}  
We compare our model with the following cross-domain recommendation baselines in the cold-start scenario:
For methods that only use ratings, we use \textbf{CMF} \cite{SinghG08}, \textbf{EMCDR} \cite{ManSJC17}, \textbf{SSCDR} \cite{KangHLY19}, \textbf{LACDR} \cite{WangZZWZH21} and \textbf{PTUPCDR} \cite{ZhuTLZXZLH22}. 
For diffusion-based methods, we use \textbf{DiffCDR} \cite{abs-2402-02182}. 
For modality-based methods, we use \textbf{CATN} \cite{ZhaoLXDS20} and \textbf{OmniMatch} \cite{DaiEAT25}.
To ensure a fair comparison, we remove the multimodal components in \modelName~and replace the user and item feature vectors with those used in the baselines (excluding modality-based baselines).
The corresponding results can be found in the ablation study under the "w/o MLLM" section.

\subsubsection{Hyper-parameters Setting.}
We conducted our experiments on an RTX 4090 GPU using PyTorch 1.11.0, Python 3.8, Ubuntu 20.04, and CUDA 11.3.
For data splitting, we follow a strategy similar to that described in \cite{ZhuTLZXZLH22}. We randomly select a portion of overlapping users as test users, with the remaining overlapping users serving as training users. We set the proportion of test users, $\beta$, at three levels of cold-start severity: 20\%, 50\%, and 80\% of all overlapping users. We employ a grid search approach to select hyperparameters. We use the Adam optimizer with learning rates set at \{0.00001, 0.0001, 0.001\}, and adjust the number of epochs between 50 and 100. The dimensionality of the feature vectors is 32. For diffusion model training, the step length $T$ during sampling is set at \{2, 5, 10, 20, 50\}. The hidden layer size of the feature extraction module is 128. The dropout rates for the feature extraction and prediction modules are \{0.1, 0.2, 0.3, 0.4, 0.5\}. The diffusion model reconstruction loss weight $\lambda$ is set at \{0.2, 0.3, 0.4, 0.5, 0.6\}. The variance values $\beta_t$ are linearly interpolated from $\beta_{start} = 1e-4$ to $\beta_{end} = 0.02$.

\begin{table}[h]
\centering
\caption{Performance of ablation study.}
\begin{tabular}{c|c|c|ccc|c}
\toprule
 Tasks & $\beta$ & Metric & \makecell{w/o \\ MLLM} & \makecell{w/o \\ diffusion} & \makecell{w/o \\ side} & \modelName\\
\midrule
\multirow{6}{*}{\makecell[c]{Task 1}}& \multirow{2}{*}{20\%} 
& MAE & 0.7010& 0.7750& 0.6990& \textbf{0.6636}\\
 &  & RMSE & 0.9482& 0.9717& 0.9291& \textbf{0.9243}
\\
\cline{2-7}
& \multirow{2}{*}{50\%} 
& MAE & 0.7070& 0.7071& 0.6915& \textbf{0.6623}
\\
 &  & RMSE & 0.9626& 0.9696& 0.9297& \textbf{0.9228}
\\
\cline{2-7}
& \multirow{2}{*}{80\%} 
& MAE & 0.6974& 0.7282& 0.7237& \textbf{0.6901}
\\
 &  & RMSE & 0.9432& 0.9834& 0.9653& \textbf{0.9570}
\\
\hline
\multirow{6}{*}{\makecell[c]{Task 2}}& \multirow{2}{*}{20\%} 
& MAE & 0.8372& 0.7777& 0.7841& \textbf{0.7646}
\\
 &  & RMSE & 1.0995& 1.0275& 1.0223& \textbf{1.0136}\\
\cline{2-7}
& \multirow{2}{*}{50\%} 
& MAE & 0.8508& 0.7987& 0.8019& \textbf{0.7889}
\\
 &  & RMSE & 1.1039& 1.0893& 1.0531& \textbf{1.0450}\\
\cline{2-7}
& \multirow{2}{*}{80\%} 
& MAE & 0.8435& 0.8034& 0.8151& \textbf{0.7954}\\
 &  & RMSE & 1.1001& 1.0729& 1.0571& \textbf{1.0466}\\
\hline
 \multirow{6}{*}{\makecell[c]{Task 3}}& \multirow{2}{*}{20\%} 
& MAE & 0.7381& 0.7346& 0.6385& \textbf{0.6235}
\\
 &  & RMSE & 0.9840& 0.9131& 0.8733& \textbf{0.8654}
\\
\cline{2-7}
& \multirow{2}{*}{50\%} 
& MAE & 0.7111& 0.7258& 0.6921& \textbf{0.6696}
\\
 &  & RMSE & 0.9523& 0.9504& 0.9311& \textbf{0.9271}
\\
\cline{2-7}
 & \multirow{2}{*}{80\%} 
& MAE & 0.7014& 0.7133& 0.7464& \textbf{0.6899}\\
 &  & RMSE & 0.9539& 0.9476& 0.9898& \textbf{0.9416}\\
 \bottomrule
\end{tabular}

\label{table3}
\end{table}

\subsection{Cross-domain Recommendation Performance Results}

In Table~\ref{table2}, we compare \modelName~with eight baseline models to evaluate their effectiveness. The first six baselines are rating-based recommendation methods, while the last two are review-based. Following the approach in \cite{abs-2402-02182}, the first six baselines use the same user-item encoding.

CMF combines data from both auxiliary and target domains, employing a shared embedding space across domains. However, it does not account for the differences between embedding spaces, which is why various CDR models outperform CMF in our experiments. EMCDR, SSCDR, LACDR, and PTUPCDR use mapping functions to transfer knowledge from the auxiliary domain to the target domain, mitigating embedding space discrepancies. DiffCDR employs a diffusion model to generate target domain feature vectors for cold-start users, partially improving performance. While these methods leverage supplementary information from the auxiliary domain, they rely solely on rating data. CATN and OmniMatch incorporate review text to enhance embeddings, capturing richer semantic information and achieving significant performance gains. However, they do not utilize multimodal item data, limiting their potential. In contrast, \modelName~leverages both image data and textual reviews, further improving recommendation performance.

The results show that as \(\beta\) decreases, the proportion of overlapping users in the training set increases, improving performance across all models. However, as \(\beta\) increases, the performance margin of \modelName~over the baselines diminishes. This is because \modelName~relies on the diffusion model to align overlapping users for effective cross-domain transfer. With fewer overlapping users (i.e., higher \(\beta\)), the diffusion model's effectiveness is reduced. For instance, in Task 3 with \(\beta = 80\%\), NDCG slightly decreases. This indicates that the scarcity of overlapping users limits the model’s ability to capture full cross-domain alignment. Nonetheless, \modelName~continues to outperform all baselines in most configurations.

\subsection{Ablation Study}

To evaluate the contributions of the individual components in \modelName, we designed three ablation experiments: \textbf{w/o MLLM}, \textbf{w/o diffusion}, and \textbf{w/o side}.
\textbf{w/o MLLM} removes the multimodal large language models (MLLMs) and multimodal data, as described above.
\textbf{w/o diffusion} removes the diffusion module entirely, treating cold-start user features in the auxiliary domain as shared cross-domain features without any additional cross-domain mapping steps.
\textbf{w/o side} excludes side users, with cross-domain feature transfer learned solely from overlapping users.
The performance of these ablated models across three cross-domain tasks is summarized in Table~\ref{table3}.

Removing MLLMs (w/o MLLM) leads to a noticeable drop in performance across all metrics and tasks. This decline highlights the importance of leveraging multimodal information and the advanced representation capabilities of MLLMs for feature extraction. When comparing the rating metrics of w/o MLLM with the baseline, we observe that w/o MLLM performs better than the baselines (except modality-based baselines), even though they all use the same user and item feature vectors. This indicates the effectiveness of our cross-domain diffusion module. 

When the diffusion module is removed (w/o diffusion), performance deteriorates, showing that the diffusion module bridges the gap between auxiliary and target domains, aligning user features with the target domain distribution. The performance of w/o diffusion is slightly better than the baseline, indicating that LLMs can achieve similar results to the baseline even without cross-domain mapping, demonstrating the effectiveness of LLMs for preliminary alignment.

Excluding side users (w/o side) also results in performance degradation, but the impact varies depending on the task and $\beta$ setting. The inclusion of side users provides additional auxiliary information that enhances feature alignment and improves performance, particularly when overlapping users are limited.

\subsection{Hyperparameter Analysis}



\subsubsection{Diffusion Steps \( T \)}
 
\begin{figure}[h]
    \centering
    \includegraphics[width=0.45\textwidth]{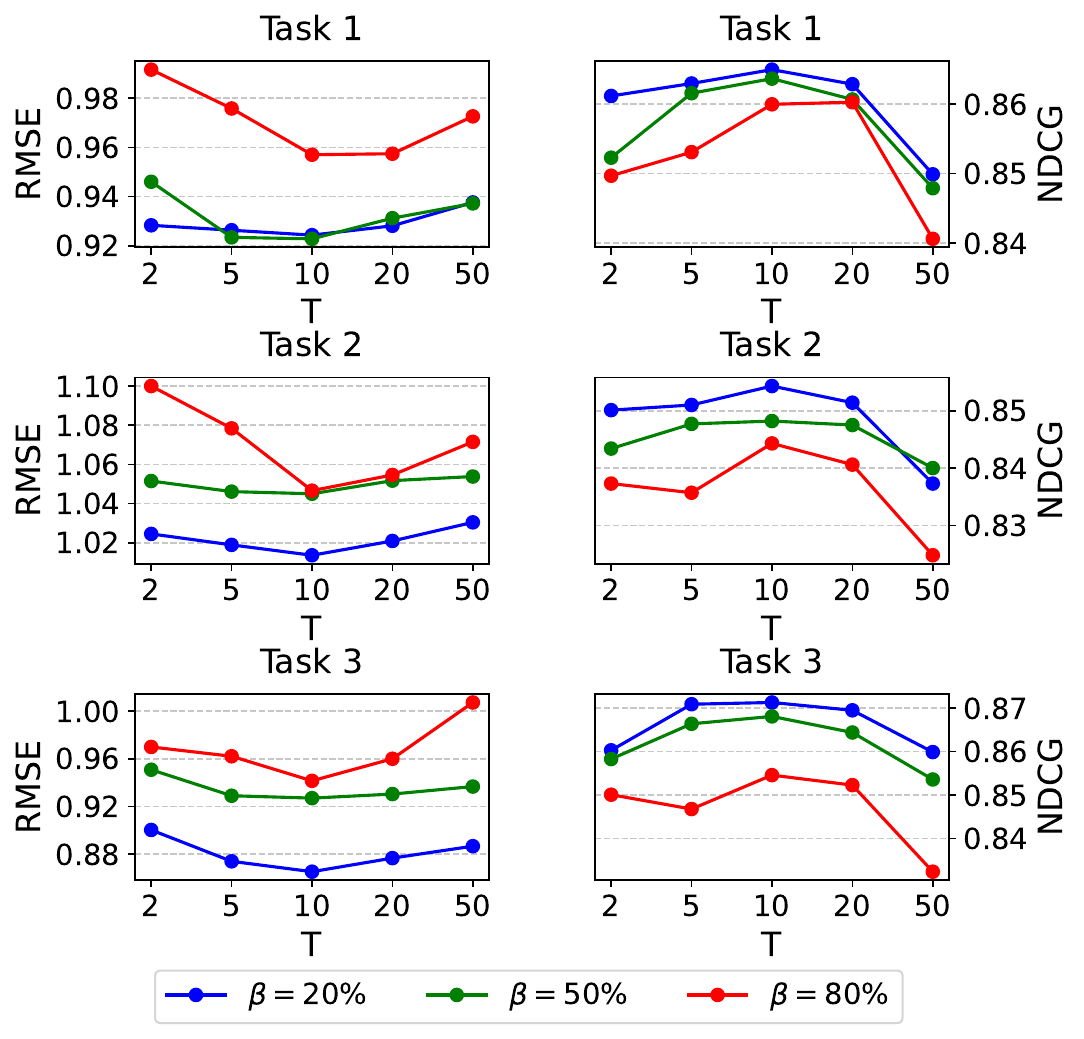}
    \caption{Impact of diffusion steps \( T \) on RMSE and NDCG for different tasks and $\beta$ values.}
    \label{fig3}
\end{figure}

We analyze the impact of the diffusion step length \(T\), a critical hyperparameter in the cross-domain diffusion module, to examine its impact on model performance. The experimental results are presented in Figures~\ref{fig3}, which illustrate the RMSE and NDCG metrics for varying \(T\) values across all three tasks and three levels of cold-start severity (\(\beta = 20\%, 50\%, 80\%\)).

\begin{figure*}[h]
    \centering
    \includegraphics[width=1\textwidth]{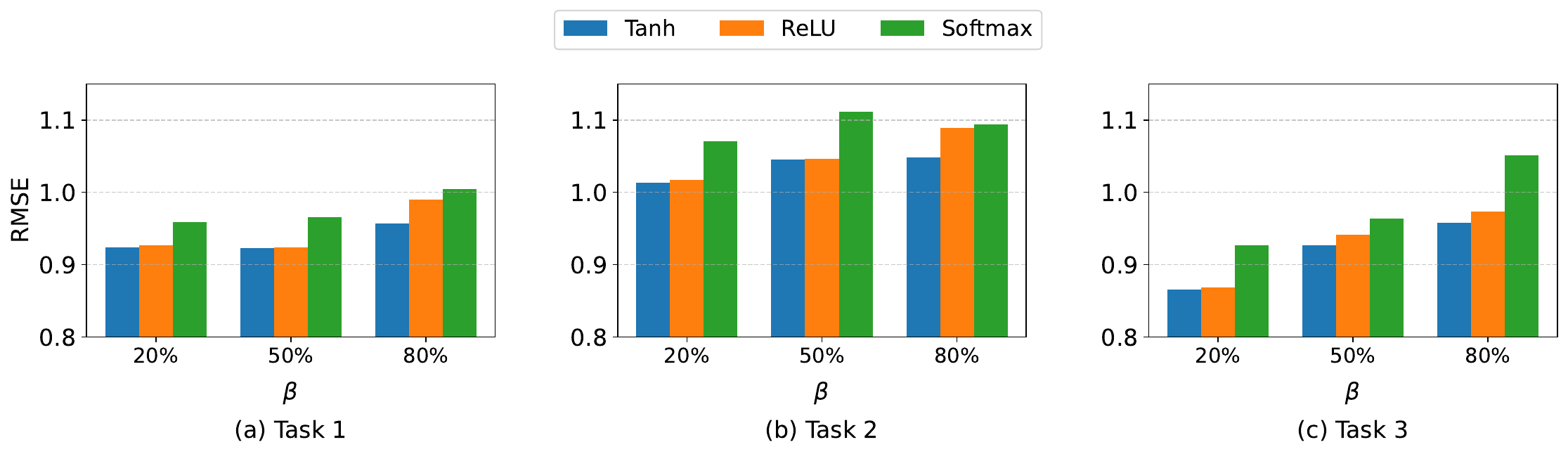}
    \caption{Parameter sensitivity analysis on activation functions.}
    \label{fig5}
\end{figure*}

On the left of Figure~\ref{fig3}, RMSE decreases with increasing \(T\) until it reaches an optimal point, then rises. This suggests that too few diffusion steps (\(T = 2\)) limit the model’s ability to reconstruct the target domain feature vectors, leading to higher RMSE values. Conversely, too large \(T\) (\(T = 50\)) introduces excessive noise into the reverse diffusion process, reducing the accuracy of the generated feature vectors.

On the right of Figure~\ref{fig3}, NDCG shows a similar trend. It improves as \(T\) increases from 2 to 10, then plateaus or declines. This confirms that a moderate \(T\) optimally balances feature alignment and noise suppression, ensuring high-quality recommendations.

\begin{table*}[htbp!]
\centering
\caption{Performance of cold-start item recommendation on RMSE and MAE.}
\begin{tabular}{c|c|ccccccccc|c}
\toprule
Tasks & Metric & CMF & EMCDR & SSCDR & LACDR  & PTUPCDR & DiffCDR & CATN    &OmniMatch& \modelName~& Improve \\
\midrule
\multirow{2}{*}{Task 1} 
& MAE & 4.3590 & 4.3635 & 4.3620 & 4.3631 & 4.3665 & 4.3635 & 4.3539 &\underline{1.1618}& \textbf{0.7977}& 31.34\%\\
 & RMSE & 4.5148 & 4.5224 & 4.5194 & 4.5194 & 4.5310 & 4.5207 & 4.4600 &\underline{1.2831}& \textbf{0.9774}& 23.83\%\\
   \hline
\multirow{2}{*}{Task 2} 
& MAE & 3.9823 & 3.9864 & 3.9868 & 3.9858 & 3.9881 & 3.9874 & 3.9877 &\underline{1.0448}& \textbf{0.9575}& 8.36\%\\
 & RMSE & 4.2040 & 4.2221 & 4.2226 & 4.2204 & 4.2264 & 4.2208 & 4.1737 &\underline{1.2260}& \textbf{1.1752}& 4.14\%\\
  \hline
\multirow{2}{*}{Task 3} 
& MAE & 4.4019 & 4.4061 & 4.4073 & 4.4074 & 4.4060 & 4.4031 & 4.3969 &\underline{0.9934}& \textbf{0.7651}& 22.98\%\\
 & RMSE & 4.5379 & 4.5690 & 4.5701 & 4.5689 & 4.5718 & 4.5600 & 4.4932 &\underline{1.1250}& \textbf{0.9358}& 16.82\%\\
\bottomrule
\end{tabular}

\label{table6}
\end{table*}

\subsubsection{Activation Function Analysis}

We analyzed the activation functions used in the feature extraction module, evaluating three commonly employed functions: \textit{tanh}, \textit{ReLU}, and \textit{softmax}. The results, shown in Figure~\ref{fig5}, report RMSE across all three tasks at varying cold-start severity levels (\(\beta = 20\%, 50\%, 80\%\)).

The \textit{tanh} activation function consistently outperforms others across all tasks and cold-start settings. This can be attributed to its ability to preserve both positive and negative values in the feature vectors, maintaining the semantic integrity of the multimodal features extracted by the LLMs. Additionally, \textit{tanh} stabilizes outputs by bounding them within the range \([-1, 1]\), which is beneficial for the training stability of the diffusion model.

From the results in Figure~\ref{fig5}, \textit{ReLU} performs worse than \textit{tanh}, particularly when the proportion of overlapping users is small. This can be attributed to two key factors: \textit{ReLU} disregards negative values, leading to information loss when processing feature vectors with both positive and negative components. Moreover, its unbounded positive output introduces instability in diffusion modeling tasks, hindering the model’s ability to learn effectively under limited training data.

As shown in Figure~\ref{fig5}, \textit{Softmax} performs the worst. This is due to its transformation of input values into a probability distribution that sums to one, distorting the original feature semantics.

\subsection{Dual Cold-Start Scenarios}
In this section, we evaluate the performance of \modelName~in the challenging dual cold-start scenario, where both users and items are cold-start. 
Cold-start items refer to items not included in the training set but appear in the test set, meaning the model has no prior interactions with these items. 

Table~\ref{table6} shows the performance of dual cold-start recommendation with \(\beta\) set at 50\%. As expected, most baselines experience a significant decline in performance due to the lack of historical data for both users and items.  
CATN uses user review data, but since cold-start items lack corresponding reviews, its performance is limited, similar to rating-based methods.  
OmniMatch also relies on review data, but unlike CATN, it generates pseudo-reviews for items to mitigate the cold-start issue. In contrast, \modelName~leverages multimodal item descriptions, including images and textual metadata, enabling accurate recommendations even for items with no interaction history and ensuring robust performance in challenging cold-start scenarios.

\subsection{Computational Cost}
\label{sec:cost}

Here we detail the computational overhead of our~\modelName~framework. 
The cost is divided into two phases. The feature extraction using MLLMs is a one-time, offline process requiring over 10 hours. In contrast, the subsequent diffusion model training is highly efficient, taking less than 30 minutes. This efficiency stems from a small number of diffusion steps ($T=10$) and operating on low-dimensional feature vectors (32-dim) instead of high-dimensional pixel data.
 
\section{Conclusion}

This paper proposes a novel cross-domain recommendation model, \modelName. The aim of \modelName~is to address two existing issues in cross-domain recommendation: 
underutilization of multimodal data which hinders cross-domain alignment and insufficient learning of the target domain’s vector space due to the neglect of side users.
In the feature extraction module, \modelName~uses an MLLM/LLM followed by an MLP to obtain feature vectors of items/users from multimodal data. The cross-domain diffusion module employs a diffusion model to learn the target domain feature distribution through side users, and to facilitate the cross-domain transfer of user feature vectors through overlapping users. This approach is then applied to cold-start users to achieve cross-domain recommendations. We conducted extensive experiments to demonstrate the effectiveness of our method and additional experiments to prove its advantages in cold-start item recommendation scenarios.

\bibliographystyle{ACM-Reference-Format}
\balance
\bibliography{acmmm}

\end{document}